\documentstyle[12pt,aasms4]{article}
\begin{document}

\title{Repeated Relativistic Ejections in GRS 1915+105}

\author{L. F. Rodr\'\i guez}
\affil{Instituto de Astronom{\'\i}a, UNAM, Apdo. Postal 70-264,
              04510 M\'exico, D.F., M\'exico}

\and

\author{I. F. Mirabel}
\affil{Service d'Astrophysique/CEA,DSM/DAPNIA, Centre d'Etudes de Saclay, 
F-91191 Gif-sur-Yvette, France and Instituto de Astronom\'\i a 
y F\'\i sica del Espacio. C.C. 67, Suc. 28. 1428, N\'u\~nez,
Buenos Aires, Argentina}

\begin{abstract}

In 1994 February-August we observed with the VLA four ejection events of 
radio emitting clouds from the high energy source GRS 1915+105. 
These events are all consistent with anti-parallel ejections of twin 
pairs of clouds moving away from the compact source at $\sim$ 0.92 
of the speed of light and angles of $\sim$ 70$^{\circ}$ with respect
to the line of sight. 
The flux ratios and time evolution of the expelled clouds are consistent 
with actual motions of the radiating matter rather than with the 
simple propagation of pulses in a medium moving at
slower velocities. The large kinetic power of the sudden, short, 
and rather discontinuous ejections exceeds by more than an order
of magnitude
the maximum steady photon luminosity of the source, suggesting that in 
GRS 1915+105 a radiation acceleration mechanism of the ejecta is 
unlikely.
As in other galactic and
extragalactic radio sources, the decrease in flux density
as a function of angular separation from the central
source shows a steepening with distance.
Additional ejection events have been observed in 1995 and 1997
and we compare them with the 1994 events.

\end{abstract}  


\section{Introduction}

Astronomical observations in the two extremes of the electromagnetic 
spectrum, in the domain of the hard X-rays on one hand, and in the domain 
of radio wavelengths on the other hand, led in recent years to the 
discovery of several galactic sources of relativistic outflows 
(see Mirabel \& Rodr\'\i guez 1995 for a review). In addition to the 
classic stellar jet source SS433 (Margon 1984), which for several years 
was believed to be unique in the Galaxy, two new jet sources that exhibit 
apparent superluminal motions were discovered: GRS 1915+105 
(Mirabel \& 
Rodr\'\i guez 1994) and GRO J1655-40 (Tingay et al. 1995; 
Hjellming \& Rupen 1995). Because of their relative proximity, this 
class of sources may offer new insights for the general understanding of 
the relativistic jets seen elsewhere in the Universe.

Due to the required strong Doppler favoritism, in distant quasars 
proper motions have been detected so far only
in the jet that is beamed toward the observer 
(Kellerman \& Owen, 1988). It should be pointed out, however, that
small-scale counterjets have been detected toward a few quasars
(e. g. Carilli et al. 1994; Taylor 1996; Walker et al. 1994)
and that future observations may detect proper motions
in these extragalactic counterjets. 
Since in the newly 
discovered galactic superluminal sources two-sided moving jets were 
detected, there was the hope that these sources would provide crucial 
information to test the physical models of relativistic jets in astrophysics. 
However, in GRO J1655-40 the jets are asymmetric and the sense of the 
asymmetry changes from event to event (Hjellming \& Rupen 1995). In this 
paper we show that the observations of repeated ejections in GRS 1915+105 
are consistent with intrinsically symmetric twin jets, which permits for 
the first time to overcome several of the fundamental uncertainties that have 
dominated the physical interpretation of superluminal motions in astronomy. 
  
\section{Observations}

The monitoring of the repeated ejections reported here were carried out with
the Very Large Array (VLA) of
NRAO\footnote{NRAO is a facility of the US 
National Science Foundation operated under
cooperative agreement by Associated Universities, Inc. }\
during 1994 February-August in the A, A/B, and B
configurations, and in 1995 August in the A configuration.
Most of the observations were made at the frequency
of 8.4 GHz (3.6-cm), using 1328+307 as  
absolute amplitude calibrator and 1923+210 as phase calibrator. 
Observations made at 20 and 2-cm on 1994 April 16 with the same absolute and
phase calibrators were used to obtain the spectra presented in
Figure 4. A 
more detailed description of the observations and first partial results from 
this monitoring were presented by Mirabel \& Rodr\'\i guez (1994) 
and Rodr\'\i guez et al. (1995).  
The data were calibrated, edited, and reduced using the
NRAO software package AIPS.

\section{Discussion}

In 1994 February-August we followed the proper motions of four pairs 
of plasma clouds moving away from the compact core of GRS 1915+105,
located at position 
$\alpha(1950) = 19^h12^m49\rlap.^s966, \delta(1950) = 
10^{\circ}51'26\rlap.{''}73$ 
(Mirabel \& Rodr\'\i guez 1994). 
An additional ejection event was observed in 1995 
August 10. The position and total flux densities
of the condensations were determined using the AIPS task
IMFIT, that fits Gaussian ellipsoids to the sources
using a least-squares criterion. In some epochs,
it was not possible to have a solution for a
weak condensation in the immediate vicinity of a bright
one, and positions and flux densities were roughly estimated
directly from the maps.
In Table 1 we list the parameters of the observations
and the angular displacements from the central core
for these four ejection events.
Figure 1 shows the source
at epoch 1994 April 09, where two condensation pairs; that ejected in 
1994 February 19 and that ejected in 1994 March 19 
are clearly visible. In Figure 2 we show the
proper motions of the condensations detected from
the four ejection events of 1994. We conclude that before and 
after the remarkable ejection event of 1994 March 19, reported 
in detail by 
Mirabel \& Rodr\'\i guez (1994), the compact source ejected other 
pairs of condensations but with flux densities one to two orders of 
magnitude fainter. 
The epoch of the ejection (as determined from the convergence of the regression 
lines on the same point of the time axis), the position angles
of the direction of motion,
and the proper motions 
for each pair are given in Table 2. The first 
three ejections of 1994
and the one observed in 1995
took place at epochs when the source was being detected by 
BATSE in the 20-100 keV energy band (Harmon et al. 1997). The time 
separation between ejections suggests a quasiperiodicity at intervals in 
the range of 20-30 days. Although the clouds appear to move always in the 
same general region of the sky, the position angles  listed in Table 1 
suggest a change  by $\sim$ 10$^{\circ}$ of the direction of ejection 
in one month. 

The four pairs moved away from the compact core with proper motions 
that, within error, are similar (see Table 2).
The proper motions away from the 
stationary core shown in Table 2 and Figure 2 
are consistent with ballistic (that is, unaccelerated) proper motions. For 
the bright pair ejected on 1994 March 19 the proper motions could be 
determined very accurately (see Table 2). In this 
case, the angular displacements were 
followed for about 50 days while the ejecta remained detectable. 
In contrast, due to limitations imposed by sensitivity and array 
configuration, only the approaching component of the pair ejected in 
1994 Jan 29 could be detected.
The proper motions listed in Table 2 for the 1995 August 10 ejection are 
unreliable since the individual condensations were resolved
only in one epoch.
In Figure 3 we show a radio map of the ejecta
pair for this event.
This event was accompanied by infrared and
X-ray activity and has been discussed in detail
by Mirabel et al. (1996).
Despite the considerable uncertainty in the parameters of the
1995 August 10 ejection, there appear to be significant
variations in proper motions and
position angle with respect to the set of ejections observed in
1994. Further research is needed to establish if
fainter events, as is the case for the 1995 ejection,
are systematically different from the type
of bright ejections seen in 1994.
Also the proper motions observed by Fender et al. (1998)
with the MERLIN interferometer for the ejection event of
1997 October (23.6$\pm$0.5 mas day$^{-1}$ for the approaching
ejecta and 10.0$\pm$0.5 mas day$^{-1}$ for the receding ejecta)
appear to be different at the 10-20 \% level from the proper motions of the
1994 events (see Table 2).
At present it is unclear if these differences are due to
an intrinsic change in the velocity of the ejecta or to
a change in the angle of ejection.

Mirabel \& Rodr\'\i guez (1994) have modeled the motions in Figure 2 
for the 1994 events 
as antiparallel ejections of twin pairs of plasma clouds moving 
at relativistic speed $\beta$ = v/c. Under this assumption,
the apparent velocities 
in the sky are given by:

$$v_{a,r} = {v~sin \theta \over {(1 \mp (v/c)~cos \theta)}} , \eqno(1)$$

\noindent the asymmetry in the proper motions between the approaching 
and receding clouds can be explained in terms of relativistic aberration. 
In the previous equation, $\theta$ is the angle
between the ejection axis and the line of sight.
At a kinematic distance of 12.5 kpc (Rodr\'\i guez et al. 1995) the proper 
motions of the approaching and receding condensations imply apparent 
velocities on the plane of the sky of $v_a$ = 1.25c and $v_r$ = 0.65c for 
the approaching and receding components respectively. The ejecta move 
with a true speed of $v$ = 0.92c at an angle $\theta$ = 70$^{\circ}$ 
with respect to 
the line of sight (Mirabel \& Rodr\'\i guez 1994). 


\subsection {Doppler boosting}

Due to relativistic aberration the brightness ratio of the approaching 
and receding condensations (measured at equal distances from the core) is 
given (Pearson \& Zensus 1987) by

$$ {S_a \over S_r} = \biggl({1 + \beta~cos \theta
\over 1 - \beta~cos \theta} \biggr)^{k-\alpha}, \eqno(2)$$

In this equation, $\alpha$ is the observed spectral index of the radiation
and $k$ is a parameter related to the geometry of the jet
($k$=2 for a continuous jet, while $k$=3 for discrete condensations).  
The flux densities of the bright condensations
ejected on 1994 March 19 are
shown in Figure 4 for the epoch 1994 April 16.
The spectral indices of both condensations are identical within
error, $\alpha$ = $-$0.8$\pm$0.1. This is consistent
with the theoretical expectations. Indeed, since the ratio
$I_\nu/\nu^3$ (with $I_\nu$ being the
intensity and $\nu$ the frequency of the radiation) is Lorentz invariant 
(Rybicki \& Lightman 1979) it can be shown that
for radiation with a power law 
spectrum, the spectral index will be the same for all observers
in moving frames of reference.
Then, since $\beta$ = 0.92 and $\theta$ = 70$^{\circ}$, 
the ratio of 
the apparent surface brightnesses for a given angular distance from the 
ejection center is predicted to be in between 6 (k = 2 for continuous jets)
and 12 (k = 3 for discrete clouds). 

Figures 5 and 6 show that for a given angular separation, the observed flux 
ratio between the approaching and receding condensations for the ejections 
of 1994 March 19 and 1994 April 21 was 8 $\pm$ 1. 
Therefore, the 
apparent motions observed in GRS 1915+105 are consistent with true bulk 
relativistic motions of the matter that emits the radio waves. 
Bodo \& Ghisellini (1995) have proposed that there could be a small contribution
of wave propagation in the pattern motions, but that most of the observed displacements
are true bulk plasma velocities.

\subsection {Fading of the ejected clouds}

Figures 5 and 6 show that for the ejections of 
1994 March 19 and April 21 the clouds faded out as they moved away 
from the core with power law dependences given
by S$_{\nu}$ $\propto$ $\phi$$^{-1.3\pm0.2}$ 
(for the 1994 March 19 ejecta)
and S$_{\nu}$ $\propto$ $\phi$$^{-2.6\pm0.5}$
(for the 1994 April 21 ejecta), with $\phi$ being
the angular displacement from the core. 
These power law indices are an average of the power
laws for the approaching and receding condensations.
For the 1994 April 21 ejecta (see Figure 6) there is some
indication that the power law indices for the 
approaching and receding condensations are different.
However, their actual values ($-$2.1$\pm$0.7 and $-$3.1$\pm$0.4,
respectively)
are not significantly different within error
and we adopt the average value given above.

The 1994 March 19 clouds were detected only within 
0.8 arc sec from the core, while the 1994 April 21 clouds could be studied
only beyond 1 arc sec from the core. This limitation comes mostly from
the fact that the VLA changed configuration and thus angular resolution 
(from 0.2 arc sec to 0.6 arc sec) during our observations
and the fainter 1994 April 21 clouds could be studied individually
only after they separated considerably.
Although we are comparing clouds ejected in two different
events, the different indices suggest a faster
decrease in flux densities beyond around 1 arc sec.
Indeed, even the bright condensations of the 1994 March 19
ejection became undetectable beyond about one arc sec from the core, while
extrapolation from their decrease as a function of angular displacement
for observations within one arc sec from the core
implied that they should had remained detectable within a few arc sec
from the core.
This steepening of the decrease in flux density
with angular separation
was first observed and discussed in detail in the context of X-ray binaries
in the case of SS~433 (Hjellming \& Johnston 1988). These authors
propose that the steepening occurs when
the jet goes from a slowed or constrained expansion close
to the central source to a free expansion regime at some distance.
Remarkably, in both GRS~1915+105 and SS~433 the decrease
close to the source can be described approximately
with $S_\nu \propto t^{-1.3}$,
while beyond a distance of $\sim 2 \times 10^{17}$ cm, $S_\nu \propto t^{-2.6}$
is observed. 

This steepening of the decrease in flux density
with angular separation
is also probably related to the similar tendency observed in the jets of some
radio galaxies, where the intensity declines as $I_\nu \propto \phi^{-x}$,
with $x$ = 1.2-1.6 in the inner regions and 
$x \sim$4 in the outer regions of the jet (Bridle \& Perley 1984). 

In Figure 7 we show the time evolution of the
flux densities for the approaching and receding 
clouds of the 1994 March 19 event. The smooth fading of the expelled 
clouds that is apparent in Figure 5 as a function of angular distance, and 
in Figure 7 as a function of time, provide additional support for the 
interpretation of the observed motions in terms of bulk motions rather 
than the propagation of pulses. In the latter, due to the inhomogeneity 
of the interstellar medium one would expect more erratic fluctuations in 
the flux density fading and in the proper motions.
Since we are dealing with ballistic motions, the angular displacements are
proportional to time, and the decrease in flux density with time shown in Figure 7
is consistent with
S$_{\nu}$ $\propto$ t$^{-1.3\pm0.2}$, that is, the same power-law index found for
the decrease in flux density with angular displacement.
This is as expected, since given the ballistic nature of the motions angular 
displacement and time are related linearly.

Are these observed power-law
dependences modified by the motion of the condensations?
If the flux density (or another parameter)
of a constant-velocity 
condensation varies with time as a power law in the frame of reference of the condensation,
the flux density (or another parameter), as seen by a stationary observer
will also vary as a power law, and furthermore,
the exponent of the power law will be the same for both frames of
reference (in other words, the exponents of these
power law dependences with time are Lorentz invariant). To verify this, assume 
that in the reference frame of the moving condensation  
the parameter varies as $\tau^{-a}$, where $\tau$ is the
time in this reference frame. Then, since $t=\tau \gamma (1 - \beta cos(\theta))$,
with $t$ being the time in the observer's frame, we find that
here the parameter will vary as $t^{-a}$, as seen by the
observer at rest. Here $\beta = v/c$ and 
$\gamma = (1 - \beta^2)^{1/2}$.

\subsection {Mass and energy of the ejecta}

To estimate the mass and energy of the clouds ejected from GRS 1915+105 
we now concentrate on the major event of 1994 March 19, for which we have 
the largest wavelength coverage and the best signal-to-noise data. We 
first calculate the magnetic fields. For the approaching condensation 
on 1994 March 24 the deconvolved
angular dimensions were of 60 $\times$ 20 mas 
(see Mirabel \& Rodr\'\i guez, 1994), with the major axis 
approximately aligned along the outflow axis. The flux density was 
655 mJy at 3.6-cm. The spectral index between 1.4 and 15 GHz was about -0.4. 

These parameters have to be corrected for relativistic effects
on the Doppler boosting and the apparent angular size. 
The ratio of the observed flux densities $S_a$ for the approaching and 
$S_r$ for the receding ejecta relative to the emitted 
(in the frame of reference of the condensation) flux density $S_o$ is
\vskip .1in

$$ {S_{a,r} \over S_o} = \delta_{a,r}^{k-\alpha}, \eqno(3)$$

where $\delta_{a,r}$ is the Doppler factor $\nu_{a,r} \over \nu_o$ for 
the approaching and receding condensations
\vskip .1in

$$ \delta_{r,a}  =  \gamma^{-1} (1 \pm \beta~cos \theta)^{-1}, \eqno(4)$$
\vskip .1in

Since $\beta$ = 0.92 and $\theta$ = 70$^{\circ}$, we get $\gamma$ = 2.6, 
and $\delta_a$ = 0.56. Using equation (3) with k = 3, the estimated flux 
density $S_o$  (in the frame of reference of the blob) at 3.6-cm
is about 
4.4 Jy (about 7 times larger than the observed flux density).

The real major axis, $L_o$, is related to the observed (in projection)
major axis, $L_a$, by
\vskip .1in

$$ {L_a \over L_o} = sin \theta~ \delta_a, \eqno(5)$$
\vskip .1in

That is, $L_o$ = 1.9 $L_a$. Then, the true dimensions of the jet are  
110 $\times$ 20 mas. We then take as characteristic dimension a geometric 
mean of
35 mas (or $7 \times 10^{15}$ cm at a distance of 12.5 kpc).

Using the formulation of Pacholczyk (1970) for minimum energy and 
integrating from 1.4 GHz to 15 GHz (the range over 
which Rodr\'\i guez et al. 1995 
measured the flux densities on that date), 
{\it and assuming only relativistic electrons,} a magnetic field of 
about 50 mGauss
and an energy of about $2 \times 10^{43}$ ergs in the relativistic
electrons are obtained. The timescale for significant
synchrotron losses is of order 50 years.
Due to the expansion of the condensations, the magnetic field
decreases with distance. A similar calculation for
the approaching condensation on 1994 April 16 gives a magnetic field of
about 7 mGauss
and an energy of about $3 \times 10^{43}$ ergs in the relativistic
electrons. As expected, the magnetic field shows a significant decrease
while the energy in relativistic electrons remains roughly constant.

Multiplying the energy by two to account also for 
the receding condensation one obtains $\sim 4 \times 10^{43}$ ergs in relativistic
electrons for that epoch. This is, of course, the {\it internal} 
relativistic energy in the radio condensations. In addition, the plasma 
clouds have a kinetic energy due to their bulk motion at 0.92c.

To estimate the mass of the blobs one can use the usual 
formula for critical frequency for synchrotron radiation
\vskip .1in

$$\nu_c(GHz) = 4.2~(\gamma/1000)^2~B(mGauss), \eqno(6)$$

\noindent
and since we saw strong radiation up to 240 GHz (IRAM observations 
made on 1993 December 6 by Rodr\'\i guez et al. 1995) with a magnetic field of 
tens of mGauss, we adopt $\gamma_i$ = 1000 as a rough average value 
for the (internal) motion of the relativistic 
electrons. Then, dividing the total energy over 
$\gamma_i m_e c^2$, where $m_e$ is the electron rest mass, one 
can obtain the number n of (relativistic) electrons in the blobs. 

The mass of the plasma clouds can be obtained in two different ways. 
1) Assuming  that there is one (non-relativistic) proton per 
(relativistic) electron one gets a proton mass estimate in the order 
of 10$^{23}$ g. 2) Assuming that there are no protons in the plasma 
clouds one obtains a similar mass because the electrons are moving so 
fast that their relativistic
mass (as opposed to their rest mass) is comparable to that of the
non-relativistic protons. In other words, the energy 
estimate in relativistic electrons equals 
$n \gamma_i m_e c^2$. On the other hand, the relativistic 
mass of the blob 
(from electrons alone) is
$n \gamma_i m_e$. Consequently, taking the relativistic energy and
dividing over $c^2$ gives the relativistic mass in electrons,
which also is about 10$^{23}$ grams. This is the mass one
has to use to estimate the relativistic kinetic energy of the 
blobs (due to their bulk motion at 0.92c or $\gamma_b$ = 2.6). 
Therefore, the presence or absence of non-relativistic 
protons affects the mass estimate by a factor of 2 only. 
Since it is likely that the condensations contain
also non-relativistic particles, the mass estimated should be
considered as a lower limit.
Then, the blobs have comparable amounts of internal 
and kinetic bulk energy. The internal energy is 
$E_i = n \gamma_i m_e c^2$ while the kinetic bulk 
energy is of order $E_b = (\gamma_b - 1) E_i$. 
Since  $\gamma_b$ = 2.6, both energies are comparable.

Assuming that there is one non-relativistic proton for each
relativistic electron,
the estimated total rest mass of the condensations is of order $\geq$10$^{23}$ g.
This mass 
could be a significant fraction of 
the mass of the inner parts of the
accretion disk, since it is equivalent to the total mass 
that would be accumulated in the disk
over several days at the typical rate 
of 10$^{-8}$ M$_{\odot}$ yr$^{-1}$ that is required to account for the 
X-ray light curve of GRS 1915+105. 
Analysis of the X-ray spectra of GRS~1915+105 suggests that 
some of the large amplitude variations in the light 
curves can be described as resulting from the rapid removal and
replenishment of matter from the inner part of an accretion disk (Belloni
et al. 1997; Mirabel et al. 1998). In fact, the 
ejection on 1995 August 10 occurred when a sharp decay and follow-up rise 
was observed in the BATSE light curve for the 20-100 keV energy band 
(Harmon et al. 1997).  

The kinetic energy of the plasma clouds suggests an acceleration 
mechanism with very large power. From the radio monitoring with 
the Nan\c{c}ay radiotelescope  it is known that the ejection event 
of 1994 March 19 lasted $\leq$ 12 hours, (Rodr\'\i guez et al. 1995), 
requiring a {\it minimum} power of the order of 10$^{40}$ erg s$^{-1}$. 
The 10$^{40}$ erg s$^{-1}$
is larger than the maximum steady photon luminosity of 
GRS 1915+105, $\sim 3 \times 10^{38}$ erg s$^{-1}$,
suggesting that a radiative acceleration mechanism is improbable.
The mechanism that accelerates and collimates the GRS~1915+105 ejecta
is yet unclear.

\section {Conclusions}

From the VLA observations of repeated ejections in 1994 January-August 
we conclude that:

1) The ballistic motions observed in GRS 1915+105 
are consistent with the hypothesis of anti-parallel ejections of 
twin pairs of condensations moving at speeds of 0.92c.

2) In 1994 January-August the ejeccions took place with a quasi-periodicity 
of 20-30 days. A change by $\sim$ 10$^{\circ}$ of the ejection axis 
in an interval of $\sim$ 1 month was observed.
However, the precession over intervals of more than one year 
is unlikely to be greater 
than 10$^{\circ}$.

3) The flux ratio and flux time evolution of the clouds moving in opposite 
directions are consistent with actual bulk motions at 
relativistic speeds of the sources of radiation. Therefore, it is unlikely 
that in GRS 1915+105 the observed motions could represent the 
propagation of pulses and/or echos.

4) The flux of the clouds expelled on 1994 March 19 appear to decrease 
slower than predicted from adiabatic expansion, as it is also
observed in other galactic and extragalactic jets.
It is possible that part of the bulk energy of the jets is
converted to magnetic flux and relativistic particles 
through dissipative interactions
with the surrounding medium or within the jet itself. 
There is a steepening of the decrease in flux density
with angular separation. Remarkably, in both GRS~1915+105 and SS~433 the decrease
close to the source can be described with $S_\nu \propto t^{-1.3}$,
while beyond a distance of $\sim 2 \times 10^{17}$ cm, $S_\nu \propto t^{-2.6}$
is observed.

5) The repeated ejecta discussed here
can reach relativistic equivalent masses of about 10$^{23}$ g 
and kinetic energies in the order of 10$^{43}$ ergs, as in the case
of the 1994 March 19 event. However, the other ejecta
events
have masses and energies an order of magnitude smaller.

6) There are ejections with kinetic power $\geq$ 10$^{40}$ erg s$^{-1}$, 
which is more than an order of magnitude larger than
the mean X-ray luminosity of 
the source. 

7) A comparison of the parameters of the 1994 ejections with those
observed in the 1997 ejection suggests the presence of differences at the
10-20 \% level in the proper motions.
At present it is unclear if these differences are due to
an intrinsic change in the velocity of the ejecta or to
a change in the angle of ejection.

\acknowledgments

We thank B.A. Harmon for information on the BATSE data.
LFR acknowledges support from DGAPA, UNAM and CONACyT, Mexico. 

\clearpage

\begin{deluxetable}{llllll}
\footnotesize
\tablecaption{EPOCH AND ANGULAR DISPLACEMENT OF 1994 EJECTA EVENTS \label{tbl-1}}
\scriptsize
\tablehead{
Epoch of & Epoch of Observation & Angular & RMS Noise & Displacement of & Displacement of  \nl
Ejection & (Days after & Resolution$^a$ & ($\mu$Jy) & Approaching Gas$^b$ & Receding Gas$^b$ \nl
(1994) & 1994 Jan 1 at 0 UT) & & & \nl
}

\startdata

Jan 29 & 56.60  & 0.25$\times$0.22; $-$50 & 36 & 0.50 & -- \nl

Jan 29 & 64.69  & 0.20$\times$0.19; 0 & 48 & 0.64 & -- \nl
 
 & & \nl

Feb 19 & 64.69  & 0.20$\times$0.19; 0  & 48 & 0.28 & -- \nl
Feb 19 & 76.59  & 0.23$\times$0.22; $-$31  & 71 & 0.49 & -- \nl
Feb 19 & 85.68  & 0.20$\times$0.19; $+$84 & 77 & 0.63 & -- \nl
Feb 19 & 92.65  & 0.22$\times$0.22; $+$71 & 36 & 0.77 & -- \nl
Feb 19 & 98.66  & 0.18$\times$0.18; $+$1 & 33 & 0.88 & 0.49 \nl
Feb 19 & 105.54 & 0.19$\times$0.18; $-$50 & 29 & 1.00 & 0.54 \nl

 & & \nl

Mar 19 & 82.62 & 0.22$\times$0.21; $+$1 & 120 & 0.08 & --  \nl
Mar 19 & 85.68  &  0.20$\times$0.19; $+$84  & 77 & 0.15 & 0.07 \nl 
Mar 19 & 92.65  &  0.22$\times$0.22; $+$71  & 36 & 0.25 & 0.13 \nl
Mar 19 & 98.66  &  0.18$\times$0.18; $+$1 & 33 & 0.38 & 0.18 \nl
Mar 19 & 105.54 &  0.19$\times$0.18; $-$50 & 29 & 0.47 & 0.25 \nl
Mar 19 & 112.52 &  0.18$\times$0.17; $-$3 & 58 & 0.62 & 0.31 \nl
Mar 19 & 119.52 &  0.21$\times$0.17; $-$42 & 40 & 0.72 & 0.37 \nl
Mar 19 & 130.50 &  0.45$\times$0.24; $+$86 & 113 & 0.93 & 0.47 \nl

 & & \nl

Apr 21 & 119.52 &  0.21$\times$0.17; $-$42  & 40 & 0.29  & -- \nl
Apr 21 & 163.27 &  0.75$\times$0.68; $-$15  & 74 & 0.97 & 0.57 \nl
Apr 21 & 170.33 &  0.74$\times$0.66; $-$13  & 63 & 1.04 & 0.63 \nl
Apr 21 & 181.29 &  0.74$\times$0.67; $-$18  & 40 & 1.31 & 0.73 \nl 
Apr 21 & 190.78 &  0.74$\times$0.67; $-$18  & 33 & 1.42 & 0.81 \nl

\enddata
\tablenotetext{a}{Angular dimensions of major$\times$minor
axis of synthesized beam in arc sec, and position angle of the
major axis in degrees.}

\tablenotetext{b}{Angular displacements from core position
given in arc sec.
Positional errors are estimated to
be $0\rlap.{''}04$, $0\rlap.{''}04$,
$0\rlap.{''}02$, and $0\rlap.{''}06$ for the 1994 Jan 29, Apr 21
Feb 19, Mar 19, and Apr 21 ejections, respectively. However, for 
the last data points observed for the 
1994 Mar 19 ejection, we use a positional error of $0\rlap.{''}06$.}

\end{deluxetable}

\clearpage

\begin{deluxetable}{llll}
\footnotesize
\tablecaption{EPOCH, POSITION ANGLE, AND PROPER MOTION OF EJECTA EVENTS \label{tbl-2}}
\scriptsize
\tablehead{
Epoch  & Position & Proper Motion of & Proper Motion of  \nl
 & Angle & Approaching Gas & Receding Gas \nl
 &  & (mas day$^{-1}$) & (mas day$^{-1}$) \nl
}
 
\startdata
 
1994 Jan 29 & 159$^\circ$$\pm$8$^\circ$ & 17$\pm$2 & --- \nl
 
1994 Feb 19 & 157$^\circ$$\pm$6$^\circ$ & 17.7$\pm$0.4 & 7$\pm$2 \nl
 
1994 Mar 19 & 149$^\circ$$\pm$4$^\circ$ & 17.5$\pm$0.3 & 9.0$\pm$0.1 \nl
 
1994 Apr 21 & 147$^\circ$$\pm$6$^\circ$ & 16.0$\pm$0.7 & 8.8$\pm$1.0 \nl
 
1995 Aug 10 & 140$^\circ$$\pm$10$^\circ$ & 11$\pm$2 & 9$\pm$2 \nl
 
\enddata
\end{deluxetable}

\clearpage

\figcaption[]{
Uniform-weight Very Large Array map of GRS~1915+105 made at 3.6-cm for the
epoch 1994 April 09. 
Contours are
-4, 4, 5, 6, 8, 10, 15, 20, 40, 60, 100, 200, 400, 800, and
1600 times 0.050 mJy beam$^{-1}$. The beam is shown in the top left corner.
The cross marks the position of GRS~1915+105. The bright pair
of condensations on each side of the source are the pair
ejected on 1994 March 19. The fainter pair of condensations ahead
of the previous pair corresponds to the 1994 February 19 ejection event.
Note also the difference of about 10$^\circ$ between the position angles
of the two pairs, suggesting precession of the jet axis. 
\label{fig-1}}

\figcaption[]{ 
Top: Angular displacements as a function of time
for four approaching condensations corresponding to
ejections that took place on (from left to right)
1994 January 29 (triangles), February 19 (squares), March 19 (circles), 
and April 21 (crosses).
Bottom: Angular displacements as a function of time
for three receding condensations corresponding to
ejections that took place on (from left to right)
1994 February 19 (squares), March 19 (circles), and April 21 
(crosses). The clouds of the 1994
January 29 ejection were relatively weak and we could not
detect unambiguously the receding component in our data. The dashed
lines are the least squares fit to the angular displacements
of the 1994 March  19 event, the brighter and better
studied. Note that the motions appear to be ballistic (that is, unaccelerated).
\label{fig-2}}

\figcaption[]{
Natural-weight Very Large Array map of GRS~1915+105 made at 3.6-cm for the
epoch 1995 August 24.
Contours are
-3, 3, 4, 6, and 8
times 0.014 mJy beam$^{-1}$. The beam is shown in the top left corner.
The cross marks the position of GRS~1915+105. This pair
of condensations is believed to have been ejected
on 1995 August 10.
\label{fig-3}}

\figcaption[]{
Flux densities at 6, 3.6, and 2-cm for the approaching and receding 
condensations of the 1994 March 19 ejection, as measured on 1994
April 16. The dashed lines are least-squares fits to the data.
Both spectral indices are consistent with $\alpha$ = --0.8$\pm$0.1.
\label{fig-4}}

\figcaption[]{
Flux densities at 3.6-cm as a function of the angular displacement 
from the core for the pairs of clouds ejected on 1994 March 19.
The solid and open circles
represent the approaching and receding components, 
respectively. The dashed lines are least-squares fits to the data.
The flux ratios of the condensations at a given angular 
displacement from the core are consistent with the expected flux ratios 
in the range of 6-12 due to Doppler-boosting of the radiation from clouds 
moving apart with relativistic bulk motions.
The decrease in flux density with angular separation from the core
is given (averaged over the approaching and
receding condensations) by S$_{\nu}$ $\propto$ $\phi$$^{-1.3\pm0.2}$. 
The flux densities are taken from Mirabel \& Rodr\'\i guez (1994), where a
statistical
error of $\sim$5\% was quoted. Here we adopt a more conservative
error of 0.1 in the log, that attempts to include possible 
systematic errors.
\label{fig-5}}

\figcaption[]{
Same as in Figure 5, but for the 
pair of clouds ejected on 
1994 April 21. 
The decrease in flux density with angular separation from the core
is given (averaged over the approaching and
receding condensations) 
by S$_{\nu}$ $\propto$ $\phi$$^{-2.6\pm0.5}$.
\label{fig-6}}

\figcaption[]{
Flux densities at 3.6-cm as a function of time for
the 1994 March 19 ejection. The solid and open circles
represent the approaching and receding components,
respectively. The dashed lines are least-squares fits to the data.
\label{fig-7}}

\clearpage

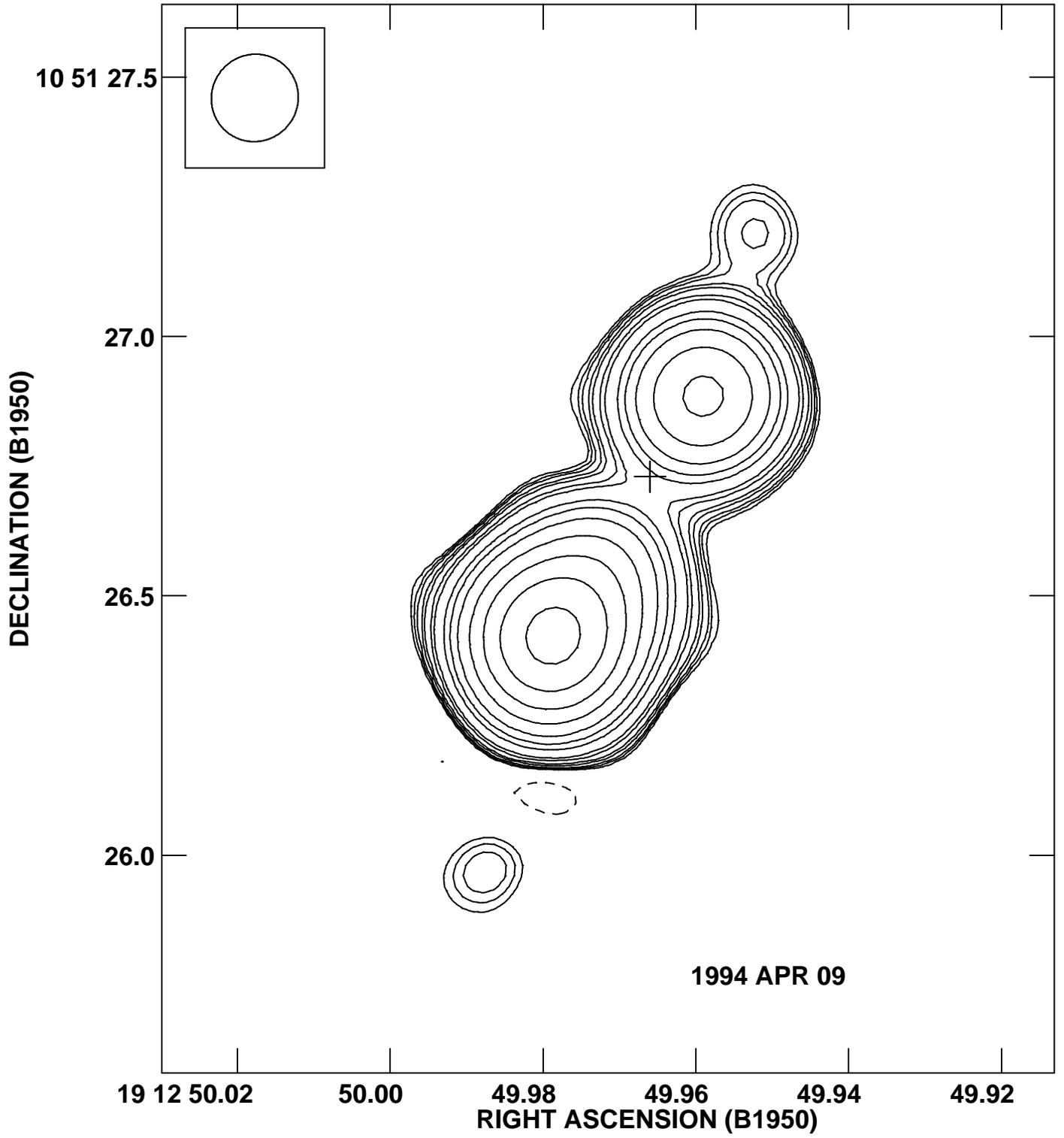
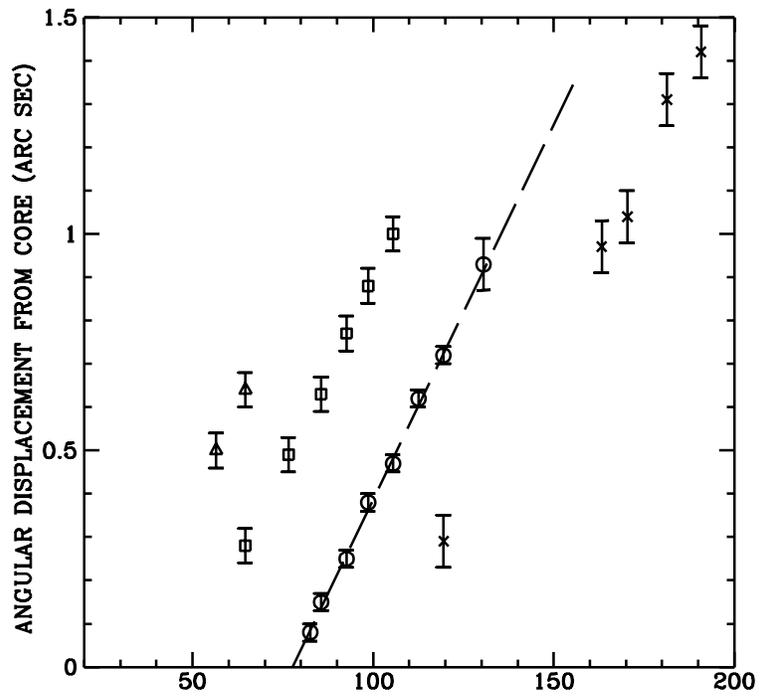
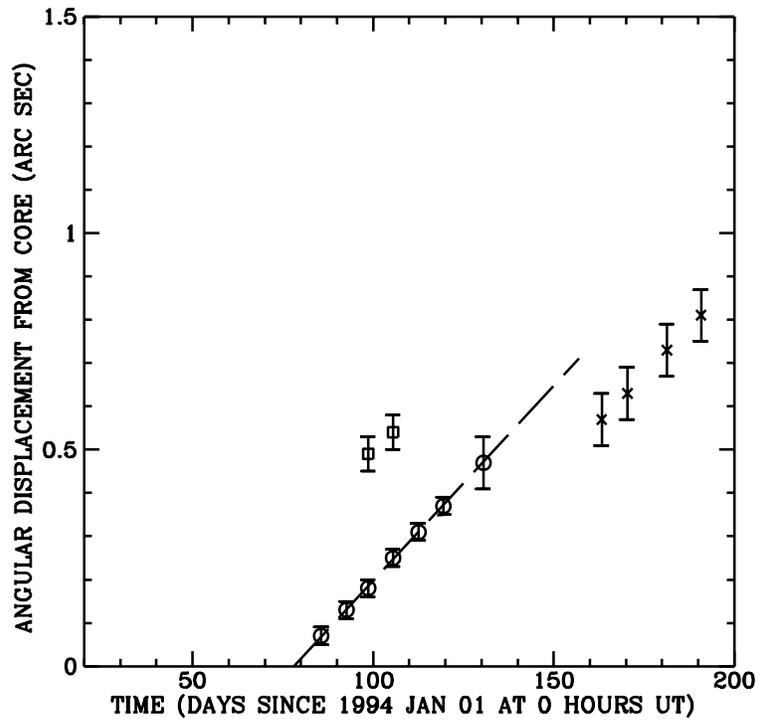
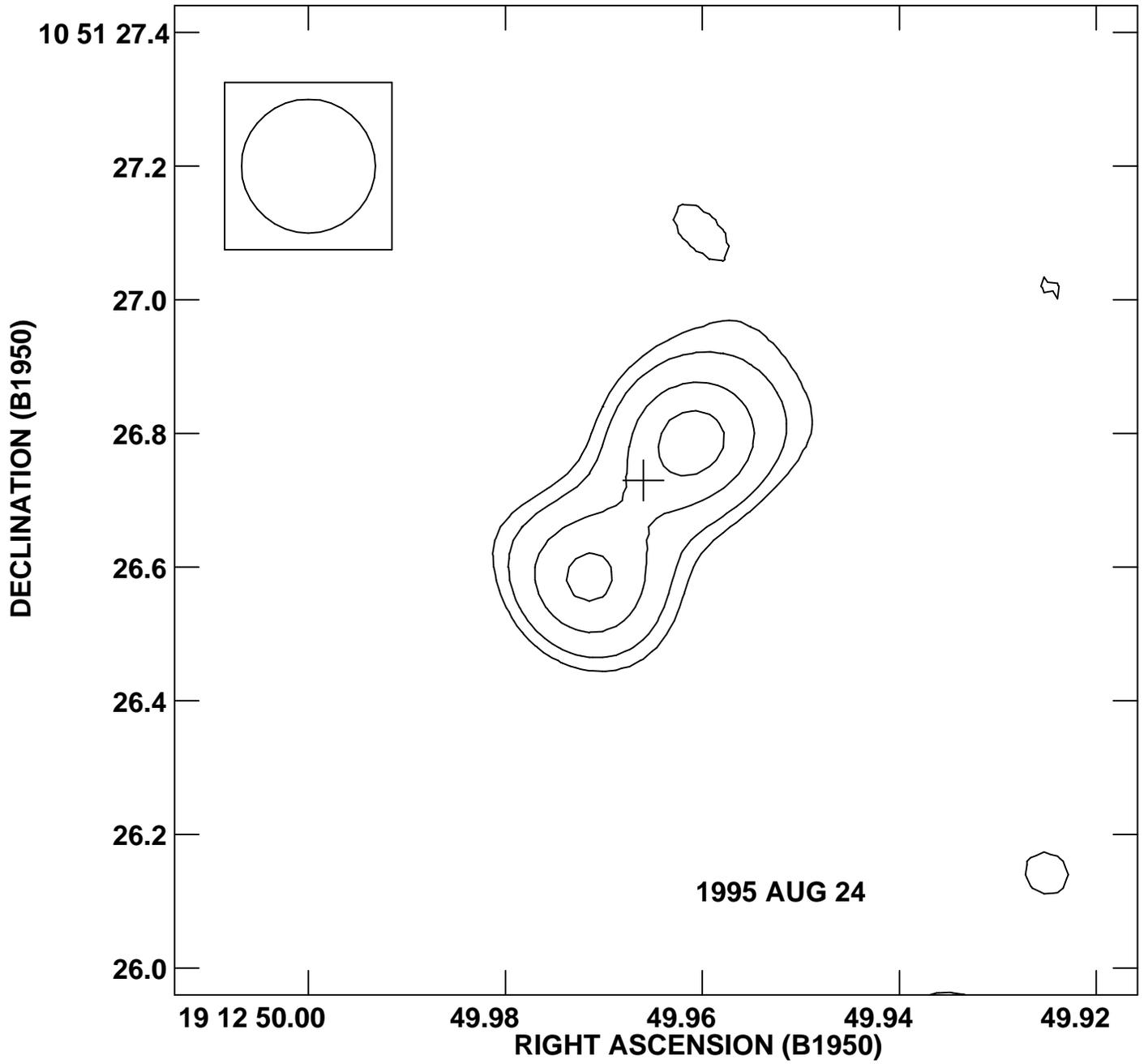
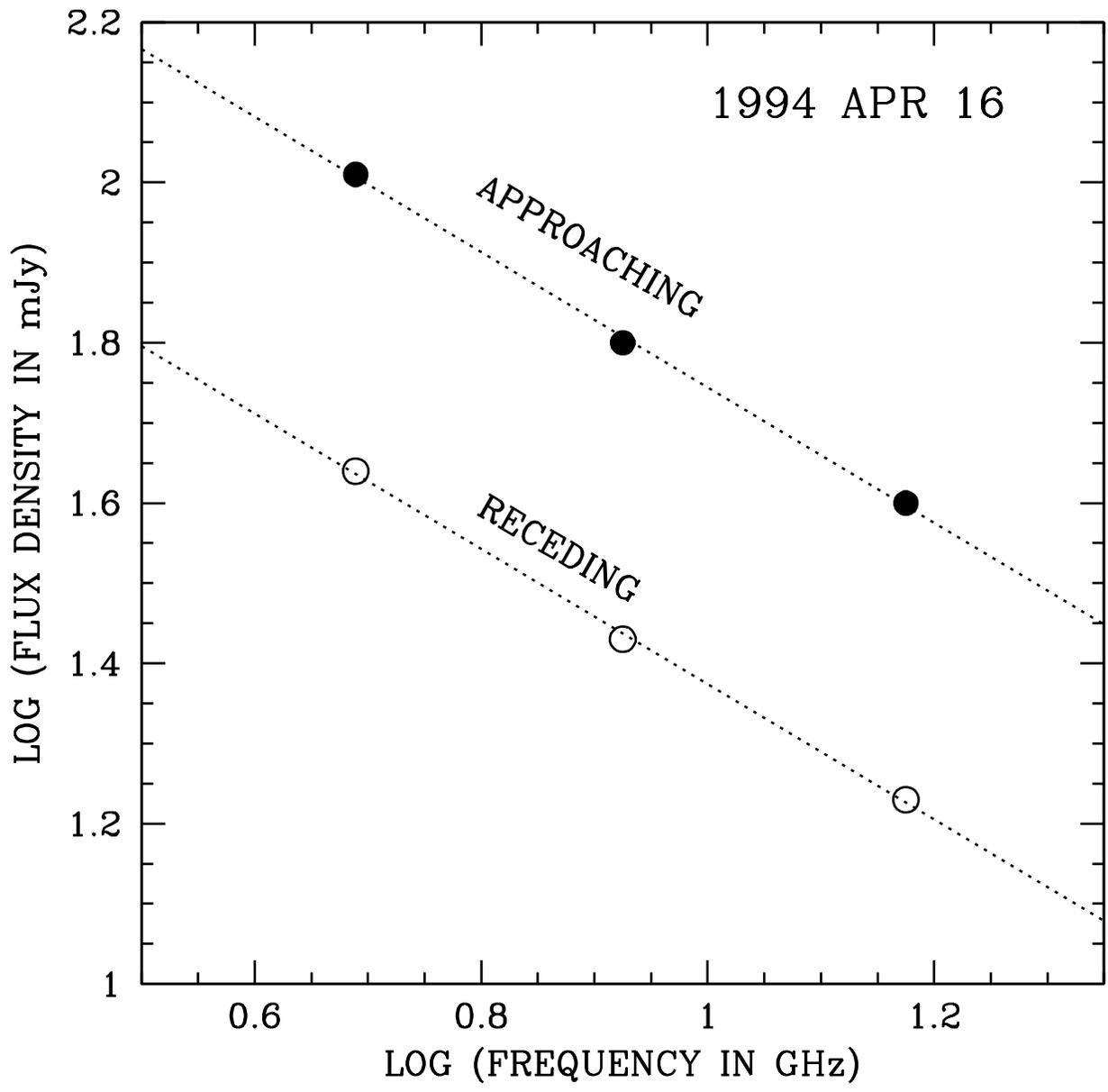
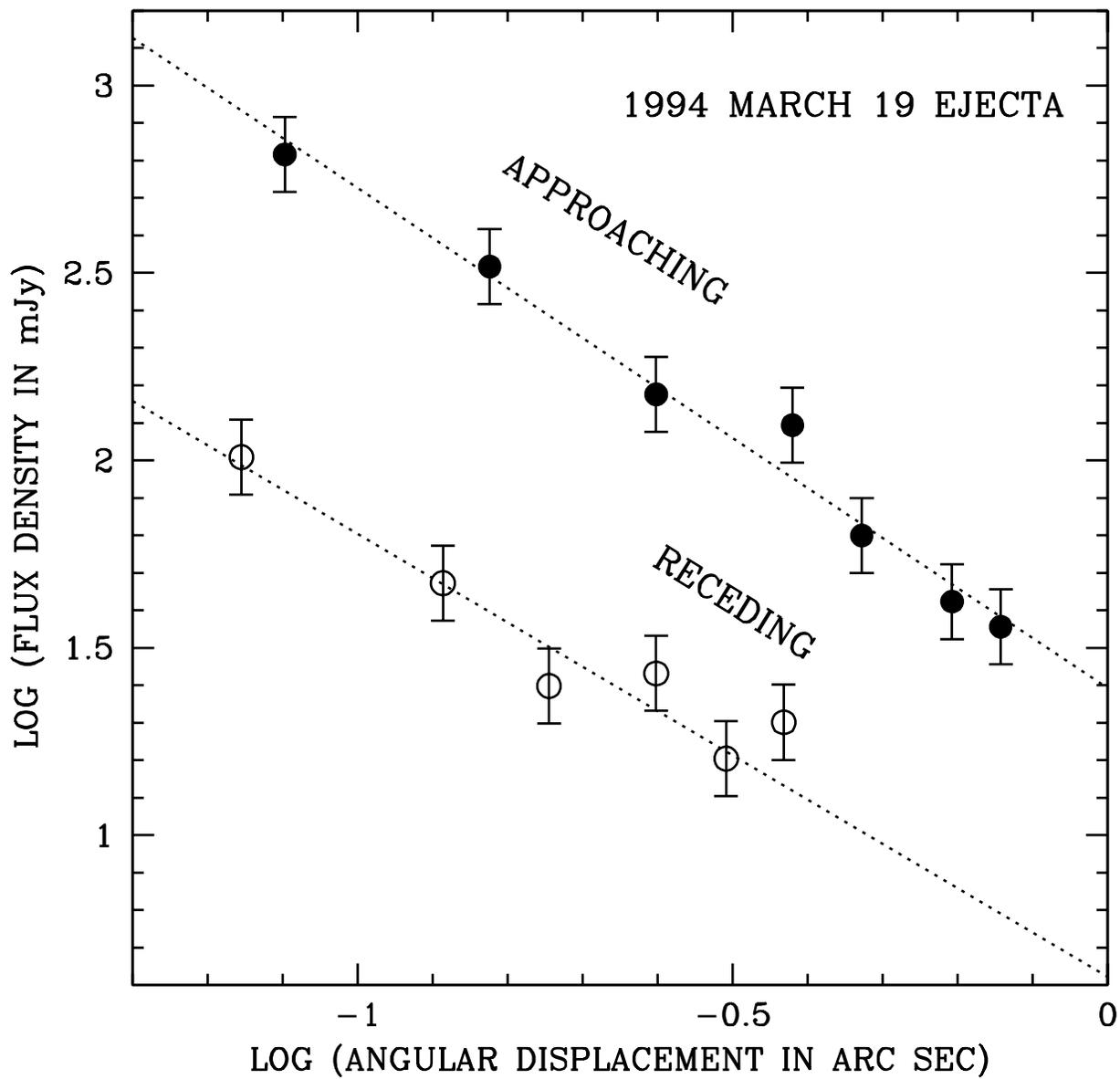
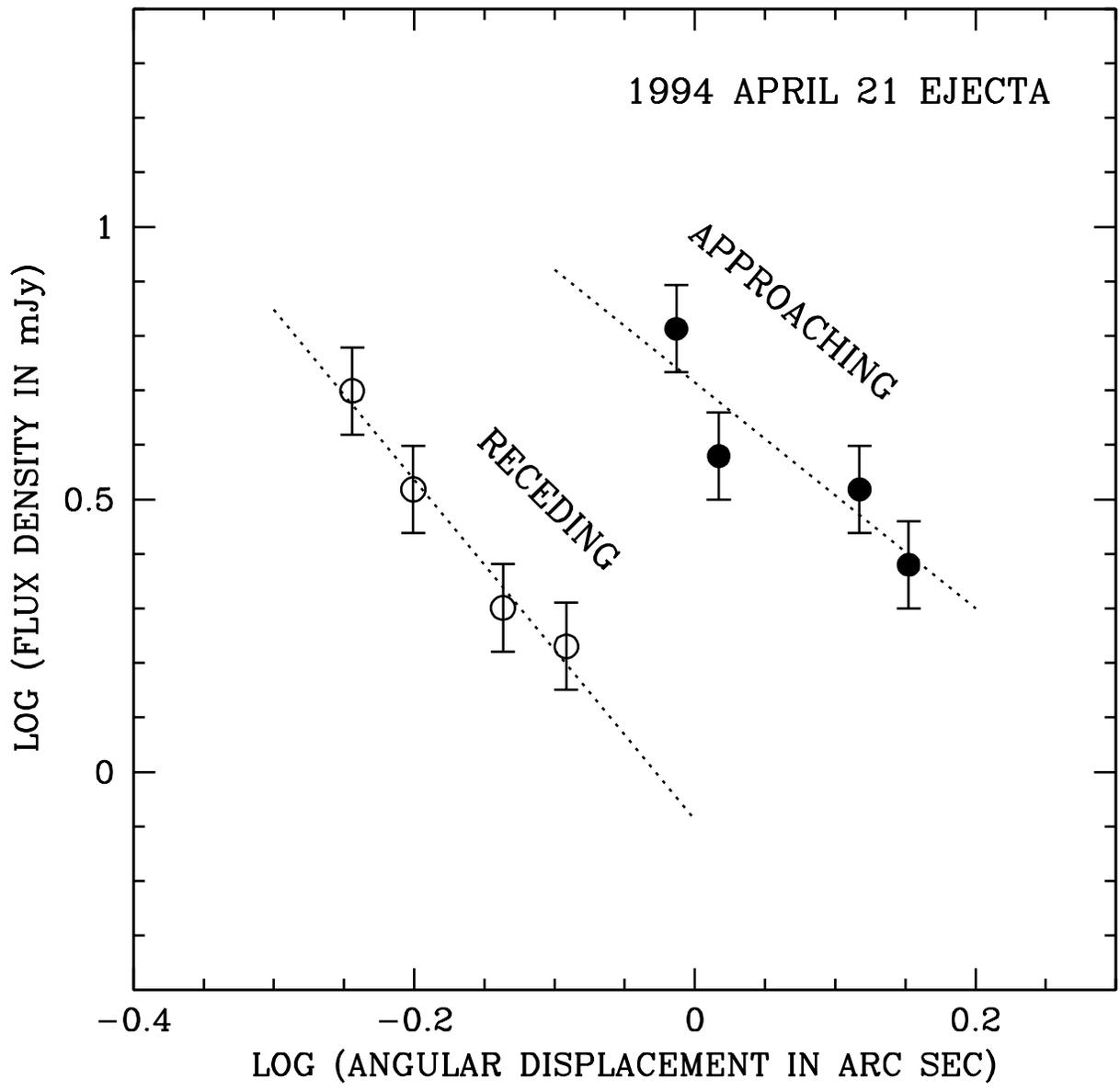
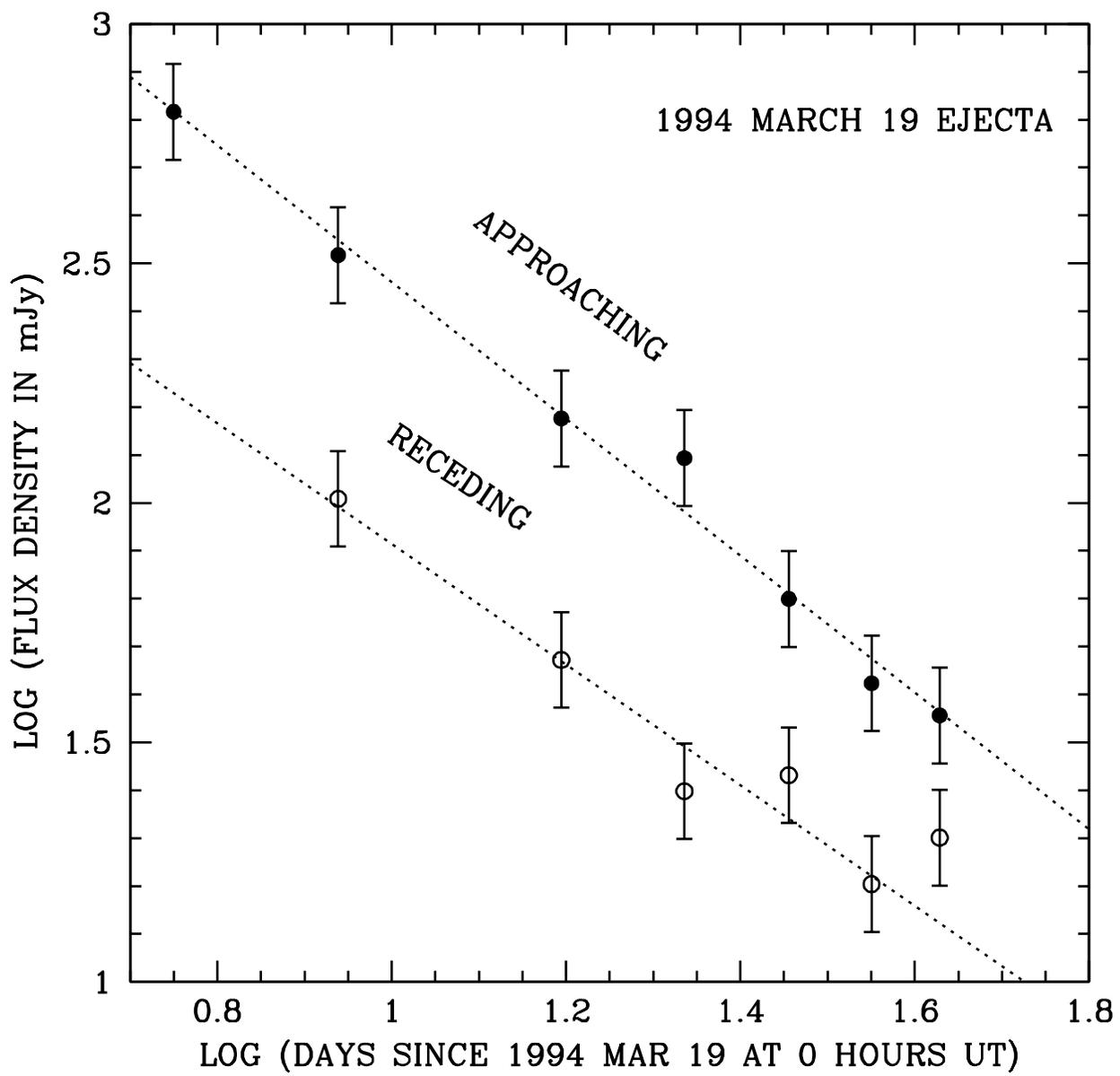
\end{document}